\documentclass[letter,journal]{IEEEtran}
\usepackage{amsmath} 
\usepackage{amsfonts}
\allowdisplaybreaks[4]
\usepackage{algorithmic}
\usepackage{algorithm}
\usepackage{array}
\usepackage{booktabs}
\usepackage[caption=false,font=normalsize,labelfont=sf,textfont=sf]{subfig}
\usepackage{textcomp}
\usepackage{color}
\usepackage{stfloats}
\usepackage{url}
\usepackage{verbatim}
\usepackage{graphicx}
\usepackage{epstopdf}
\usepackage{cite}
\newtheorem{remark}{Remark}

\begin{document}

\title{\LARGE Transmissive Reconfigurable Intelligent Surface Transmitter Empowered Cognitive RSMA Networks}

\author{Ziwei~Liu,~Wen~Chen,~\IEEEmembership{Senior Member,~IEEE,}~Zhendong~Li,~Jinhong~Yuan,~\IEEEmembership{Fellow,~IEEE},~Qingqing Wu,~\IEEEmembership{Member,~IEEE,}~and~Kunlun Wang,~\IEEEmembership{Member,~IEEE}
	\thanks{This work is supported by National key project 2020YFB1807700, NSFC 62071296, Shanghai 22JC1404000, 20JC1416502, and PKX2021-D02.}	
	\thanks{Z. Liu, W. Chen, Z. Li, and Q. Wu are with the Department of Electronic Engineering, Shanghai Jiao Tong University, Shanghai 200240, China (e-mail: ziweiliu@sjtu.edu.cn; wenchen@sjtu.edu.cn; lizhendong@sjtu.edu.cn; wu.qq1010@gmail.com).}
\thanks{J. Yuan is with the School of Electrical Engineering and Telecommunications, University of New South Wales, Sydney, NSW 2052, Australia (e-mail: j.yuan@unsw.edu.au). }	
\thanks{K. Wang is with the School of Communication and Electronic Engi-neering, East China Normal University, Shanghai 200241, China (e-mail:klwang@cee.ecnu.edu.cn).}
}

%

\maketitle

\begin{abstract}
In this paper, we investigated the downlink transmission problem of a cognitive radio network (CRN) equipped with a novel transmissive reconfigurable intelligent surface (TRIS) transmitter. In order to achieve low power consumption and high-rate multi-streams communication, time-modulated arrays (TMA) is implemented and users access the network using rate splitting multiple access (RSMA). With such a network framework, a multi-objective optimization problem with joint design of the precoding matrix and the common stream rate is constructed to achieve higher energy efficiency (EE) and spectral efficiency (SE). Since the objective function is a non-convex fractional function, we proposed a joint optimization algorithm based on difference-of-convex (DC) programming  and successive convex approximation (SCA). Numerical results show that under this framework the proposed algorithm can considerably improve and balance the EE and SE.
\end{abstract}

\section{Introduction}
\IEEEPARstart{F}{uture} wireless communications need to satisfy the demands of massive access and fast growing-data flow, which lead to increasing problems of system energy consumption, spectrum shortage, and inter-device interference\cite{8869705}. Therefore, there is an urgent need to advance the development of technologies to tackle the aforesaid problems.

 As for the spectrum shortage problem, radio cognitive network (CRN) was introduced to manage spectrum resources\cite{5783948}, where users are divided into primary users (PUs) and cognitive users (CUs). By interacting with the wireless environment and sensing the underutilized spectrum which can be dynamically allocated to the CUs without causing interference to the PUs. Therefore, the spectrum resource is fully utilized, and greatly improving the spectrum efficiency (SE) \cite{9475472}.

In order to achieve low-cost and low-energy consumption communication deployments, reflective reconfigurable intelligent surface (RIS) is being focused on\cite{9647934}. RIS consists of a large number of low-cost passive elements and is considered as a key technology for next-generation wireless communication technologies on account of its ability to improve the wireless propagation constructively. Due to this characteristic, reflective RIS is used for assisted communication\cite{9531372}. Recently, a transmissive RIS (TRIS) transmitter has been proposed \cite{9570775}, and TRIS can achieve higher aperture efficiency due to its structural features, which avoid the problem of feed source occlusion and the interference problem of reflected waves. Meanwhile, since the transmissive architecture is equipped with only one feed antenna, the energy consumption of the system is lower than that of the conventional transceiver architecture. In addition, combined with the time-modulated array, it can realize digital signal modulation of arbitrary order\cite{he} and multi-streams communication. 

Since  this novel TRIS is equipped with only one single antenna\cite{9570775}, the non-orthogonal access method is considered. Rate splitting multiple access (RSMA), as a general non-orthogonal access scheme, can achieve a higher SE and has recently attracted considerable attention and research for the multi-antenna system, which generalizes and coordinates two extreme interference management strategies \cite{9831440,9832611}. The first extreme interference management strategy fully treats the residual interference as noise, and this is the scheme applied in space division multiple access (SDMA), which utilizes linear precoding to distinguish users spatially. The second extreme interference management strategy fully decodes and cancels the multi-user interference, and this is the scheme applied in non-orthogonal multiple access (NOMA), which utilizes successive interference cancellation (SIC) techniques to cancel the interference\cite{9831440}. Combining the above two schemes, RSMA utilizes both linear precoding and SIC techniques to decode part of the interference and treat the remaining part as noise, thus can reduce the interference at the users. In addition, RIS-assisted communication system using RSMA for access is investigated \cite{9832618}\cite{9145189}, and the results show that this scheme outperforms conventional schemes in terms of SE and energy efficiency (EE). The other way around, based on the multi-streams scheme design of this paper, in order to optimize the RIS element phase, the precoding matrix optimization of RSMA is naturally considered.

Inspired by the above work, we constructed a novel TRIS transmitter architecture to achieve low power consumption and high-rate multi-streams communication and jointly optimize the parameters of RIS and RSMA to improve the EE and SE of the system. Meanwhile, we investigate the trade-off between the two. Due to coupled optimization variables and fractional functions, we propose a joint optimization algorithm based on difference-of-convex (DC) programming and successive convex approximation (SCA) to solve the non-convex problem.

\section{System model and problem formulation}
As shown in Fig. 1, we consider a downlink transmission CRN with RSMA. The primary base station (PBS) serves $N$ single-antenna PUs, and the cognitive base station (CBS) serves $K$ single-antenna CUs.

\subsection{TRIS Transmitter Characteristics and Multi-stream Scheme}
\begin{figure}
	\centerline{\includegraphics[width=8.5cm]{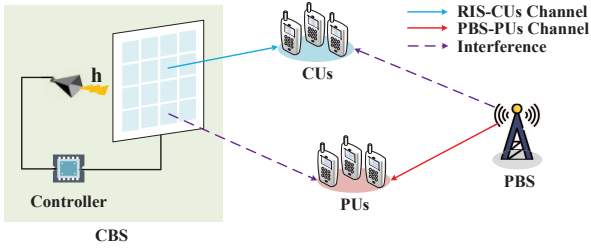}}
	\caption{TRIS transmitter empowered cognitive RSMA networks.}
\end{figure}
The CBS consists of a feed antenna, a controller, and a TRIS, and the TRIS has $M = {M_r} \times {M_c}$ transmissive elements arranged in a uniform planar array (UPA) with the centers of adjacent elements spaced at ${d_f} = {\lambda _c}/2$, where ${\lambda _c}$ is the carrier wavelength. TRIS differs from reflective or active RIS in that it serves to load information (including user and beam information) and spatial diversity, and does not additionally process the signal. Based on the above differences and characteristics, this transmitter structure achieves beamforming and multi-stream communication in four steps as described below.

 Firstly, the signal ${\bf{s}}$ is precoded in the controller through the precoding matrix ${\bf{P}}$ to form a directional complex signal ${\bf{Ps}}$. Secondly, the complex signal is modulated by time array to generate a control signal ${\rm{Crtl}}\left( t \right)$ whose duty cycle is determined by the amplitude and phase of the complex signal\cite{he}. Thirdly, the control signal controls the phases of the TRIS elements to load the information onto the TRIS. Finally, the feed antenna transmits the carrier wave to carry the loaded TRIS information to the users. 
 
For time-modulated arrays, the amplitude $A$ and phase ${\varphi _0}$ of the coded information are mapped by adjusting the 0-state digital signal start time ${t_{on}}$ and 0-state digital signal duration $\tau $ of the control signal, which can be expressed as follows
\begin{equation}
	\setlength{\abovedisplayskip}{3pt}
	\setlength{\belowdisplayskip}{3pt}
	A/{A_{\max }} = \sin \left( {\pi \tau /{T_p}} \right),
\end{equation} 
and
\begin{equation}
	\setlength{\abovedisplayskip}{3pt}
	\setlength{\belowdisplayskip}{3pt}
	- \pi \left( {2{t_{on}} + \tau } \right)/{T_p} = {\varphi _0} + 2k\pi ,\forall k,
\end{equation} 
where ${T_p}$ is the code element time and ${A_{\max }}$ is the maximum amplitude of the digitally modulated signal.

Notice that the control signal superimposes information from all users through a precoding matrix in one code element and information about different code elements of the control signal is loaded onto different TRIS elements at the same time. This way each TRIS element contains the same information for all users, but with different directivity information, and each TRIS element serves all users. Therefore, beamforming and multi-stream communication are realized.

\subsection{Channel Model} 
In the CBS, the feed antenna to the TRIS is unobstructed and the physical distance is less than the Rayleigh distance $2D/{\lambda _c}$, usually called the near-field. Based on the spherical wave assumption\cite{2005Fundamentals}, the near-field line-of-sight (LOS) channel can be derived in the following form
\begin{equation}
		\setlength{\abovedisplayskip}{3pt}
	\setlength{\belowdisplayskip}{3pt}
\begin{array}{l}
	{\bf{h}}  = \alpha \left[ {{e^{ - j2\pi {D_{FR}}\left( {1,1} \right)}}, \cdots ,{e^{ - j2\pi {D_{FR}}\left( {1,{M_c}} \right)}},} \right.\\
	{\rm{~~~~~~}}{\left. { \cdots ,{e^{ - j2\pi {D_{FR}}\left( {{M_r},1} \right)}}, \cdots ,{e^{ - j2\pi {D_{FR}}\left( {{M_r},{M_c}} \right)}}} \right]^T},
\end{array}
\end{equation}
where $\alpha $ denotes the channel gain from the feed antenna to the TRIS, and ${D_{FR}}\left( {{m_r},{m_c}} \right) = \sqrt {d_0^2 + d_{{m_r},{m_c}}^2}$ denotes the Euclidean distance from the feed antenna to the $\left( {{m_r},{m_c}} \right)$-th element of the TRIS. ${d_{0}}$ denotes the distance from the feed antenna to the TRIS center and ${d_{{m_r},{m_c}}}$ denotes the distance from the $\left( {{m_r},{m_c}} \right)$-th element of the TRIS to the center of the RIS, which can be expressed as ${d_{{m_r},{m_c}}} = {d_f}\sqrt {\Delta _r^2 + \Delta _c^2} $, ${\Delta _r} = ({2{m_r} - {M_r} - 1})/{2}$ and ${\Delta _c} = ({2{m_c} - {M_c} - 1})/{2}$ denotes the incremental index.

For the channel from TRIS to CUs (The channel model between TRIS to PUs ${{\bf{g}}_n}$ is the same as the CUs), we consider the existence of line-of-sight (LoS) part and non-line-of-sight (NLoS) part, modeled as the following Rician fading channel
\begin{equation}
		\setlength{\abovedisplayskip}{3pt}
	\setlength{\belowdisplayskip}{3pt}
	{{\bf{g}}_k} = {\xi _k}\left( {\sqrt {\frac{{{\kappa _v}}}{{{\kappa _v} + 1}}} {{\bf{g}}_{{\rm{LoS}}}} + \sqrt {\frac{1}{{{\kappa _v} + 1}}} {{\bf{g}}_{{\rm{NLoS}}}}} \right),\forall k,
\end{equation}
where ${\xi _k}$ represents the path loss, and ${\kappa _v}$ represents the Rician factor. The LoS channel can be expressed as
\begin{equation}
		\setlength{\abovedisplayskip}{3pt}
	\setlength{\belowdisplayskip}{3pt}
	{{\bf{g}}_{{\rm{LoS}}}} = {\left[ {{e^{ - j2\pi {\delta _r}{{\bf{m}}_r}}}} \right]^T} \otimes {\left[ {{e^{ - j2\pi {\delta _c}{{\bf{m}}_c}}}} \right]^T},
\end{equation}
where ${{\bf{m}}_r} = \left[ {0,1, \cdots ,{M_r} - 1} \right]$ , ${{\bf{m}}_c} = \left[ {0,1, \cdots ,{M_c} - 1} \right]$. ${\delta _r} = {{d_f}\sin \varphi \cos \psi }/{\lambda }$ , and ${\delta _c} = {{d_f}\sin \varphi \sin \psi }/{\lambda }$. $\varphi $ and $\psi $ represent the azimuth and pitch angles of the electromagnetic wave at the transmitting TRIS element. The NLoS channel obeys a circularly symmetric complex Gaussian distribution, i.e., ${{\bf{g}}_{{\rm{NLoS}}}} \sim {\cal C}{\cal N}\left( {0,{{\bf{I}}_{{M_r}{M_c}}}} \right)$. The CSI of both PUs and CUs is assumed to be known at the transmitter, and the signaling overhead for CSI acquisition can be obtained in \cite{9685625}.
\subsection{Signal Model And Transmission Scheme} 
In this paper, users apply the RSMA to access the network, where the information is divided into two separate parts, the common and the private stream. The information stream is precoded in the TRIS controller utilizing the beamformer, and all CUs share the same codebook. For the common information stream part of CUs ${s_c} \in \mathbb{C}$ is coded jointly using the coding matrix ${{\bf{p}}_c} \in {\mathbb{C}^{M \times 1}}$, and the private information stream of CUs ${s_{k}} \in \mathbb{C}$ is coded separately using the coding matrix ${{\bf{p}}_{k}} \in {\mathbb{C}^{M \times 1}}$. Then, the signal to be modulated can be expressed as
\begin{equation}
		\setlength{\abovedisplayskip}{3pt}
	\setlength{\belowdisplayskip}{3pt}
	{\bf{x}} = {\bf{Ps}},
\end{equation}
where ${\bf{P}} = \left[ {{{\bf{p}}_c},{{\bf{p}}_1},{{\bf{p}}_2}, \cdots ,{{\bf{p}}_K}} \right] \in {\mathbb{C}^{M \times \left( {K + 1} \right)}}$, and ${\bf{s}} = {\left[ {{s_c},{s_1},{s_2}, \cdots ,{s_K}} \right]^T} \in {\mathbb{C}^{\left( {K + 1} \right) \times 1}}$.

When the users decode common stream information, the private stream's interference is considered as noise. After decoding the common stream information, the common stream information is cancelled from the received signal. Then, users decode their private stream information from the received signal. From the above description, the equivalent common stream and private stream signal-to-noise ratio of the $k$-th CU can be expressed as follows
\begin{equation}
	\setlength{\abovedisplayskip}{3pt}
	\setlength{\belowdisplayskip}{3pt}
	{\gamma _{c,k}} = \frac{{{{\left| {{\bf{g}}_k^H{\rm{diag}}\left( {\bf{h}} \right){{\bf{p}}_c}} \right|}^2}}}{{\sum\nolimits_{k=1}^K {{{\left| {{\bf{g}}_k^H{\rm{diag}}\left( {\bf{h}} \right){{\bf{p}}_{k}}} \right|}^2}}  + {\sigma ^2}}},\forall k,
\end{equation}
and
\begin{equation}
	\setlength{\abovedisplayskip}{3pt}
	\setlength{\belowdisplayskip}{3pt}
	{\gamma _{p,k}} = \frac{{{{\left| {{\bf{g}}_k^H{\rm{diag}}\left( {\bf{h}} \right){{\bf{p}}_{k}}} \right|}^2}}}{{\sum\nolimits_{i \ne k}^K {{{\left| {{\bf{g}}_k^H{\rm{diag}}\left( {\bf{h}} \right){{\bf{p}}_{i}}} \right|}^2}}  + {\sigma ^2}}},\forall k.
\end{equation}
Then the corresponding achievable rate for the common and private streams can be expressed as
\begin{equation}
		\setlength{\abovedisplayskip}{3pt}
	\setlength{\belowdisplayskip}{3pt}
	{R_{i,k}} = W{\log _2}\left( {1 + {\gamma _{i,k}}} \right), i \in \left\{{c,p}\right\}, \forall k.
\end{equation}
To ensure that all CUs are able to decode the common stream information, the following constraints need to be satisfied
\begin{equation}
		\setlength{\abovedisplayskip}{3pt}
	\setlength{\belowdisplayskip}{3pt}
	{R_c} = \min \left( {{R_{c,1}},{R_{c,2}}, \cdots ,{R_{c,K}}} \right).
\end{equation}
The common stream rate is shared by all CUs, which jointly participate in the encoding of the common information stream and need to satisfy the following constraints
\begin{equation}
	\setlength{\abovedisplayskip}{3pt}
	\setlength{\belowdisplayskip}{3pt}
	\sum\nolimits_{k=1}^K {{C_k}}  = {R_c},
\end{equation}
where ${C_k}$ denotes the equivalent common stream rate of the $k$-th CU.
Combining the above common stream rate and private stream rate, the total achievable rate of CUs can be obtained as follows
\begin{equation}
		\setlength{\abovedisplayskip}{3pt}
	\setlength{\belowdisplayskip}{3pt}
	{R_{tot}} = \sum\nolimits_{k=1}^K {\left( {{C_k} + {R_{p,k}}} \right)} .
\end{equation}
 Let $l \in \left\{ {c,1,2, \cdots ,K} \right\}$. The total energy consumption of the cognitive network can be expressed as
\begin{equation}
		\setlength{\abovedisplayskip}{3pt}
	\setlength{\belowdisplayskip}{3pt}
	{P_{tot}} = {\sum\nolimits_l {\left\| {{{\bf{p}}_l}} \right\|} ^2} + {P_{cir}} \le {P_{\max }}.\label{power}
\end{equation}
where ${P_{cir}}$ denotes the circuit power consumption, and ${P_{max}}$ denotes the maximum power for CBS, which is related to the maximum transmitting power of the TRIS elements.
\subsection{Problem Formulation} In this paper, the SE and EE of the cognitive networks are maximized by optimizing the precoding matrix ${\bf{P}} = \left[ {{{\bf{p}}_c},{{\bf{p}}_1},{{\bf{p}}_2}, \cdots ,{{\bf{p}}_K}} \right]$ and the common stream rate vector ${\bf{c}} = \left[ {{C_1},{C_2}, \cdots ,{C_K}} \right]$, while guaranteeing the quality of service (QoS) of the PUs. The corresponding ${\eta _{SE}}$ and ${\eta _{EE}}$ can be expressed as ${\eta _{SE}} = {R_{tot}}/W,~{\eta _{EE}} = {R_{tot}}/{P_{tot}}.$ Then this optimization problem can be formulated as
\begin{subequations}
		\setlength{\abovedisplayskip}{4pt}
	\setlength{\belowdisplayskip}{4pt}
	\begin{align}
&({\rm{{P}}{{1}}}):~\mathop {\max }\limits_{{\bf{P}},{\bf{c}}}{\rm{~}}{\eta _{SE}},\\
            &~~~~~~~~~\mathop {\max }\limits_{{\bf{P}},{\bf{c}}}{\rm{~}}{\eta _{EE}},\\
&~{\rm{s}}{\rm{.t}}{\rm{.}}~~~~{P_{tot}} \le {P_{\max }},\label{P1C1}\\
&~~~~~~~~\sum\nolimits_{k=1}^K {{C_k}}  \le {R_c},\label{Ck1} \\
&~~~~~~~~~{C_k} \ge 0,\forall k,\label{Ck2}\\
&~~~~~~~~~{{\bf{p}}_c}\succeq0,{{\bf{p}}_k}\succeq0,\forall k,\\
&~~~~~~~~~{\left| {{\bf{g}}_n^H{\rm{diag}}\left( {\bf{h}} \right){{\bf{p}}_c}} \right|^2} \le {I_{c,th}},\forall n\label{ct}\\
&~~~~~~~~\sum\nolimits_{k=1}^K {{{\left| {{\bf{g}}_n^H{\rm{diag}}\left( {\bf{h}} \right){{\bf{p}}_{k}}} \right|}^2}}  \le {I_{p,th}},\forall n,\label{pt}\\
&~~~~~~~~~{C_k} + {R_{p,k}} \ge {R_{th}},\forall k.\label{P1C7}
	\end{align}
\end{subequations}
Note that constraints (\ref{ct})-(\ref{pt}) are required to guarantee the QoS of PUs, and thus it is required that the interference power of the CUs' common and private streams to the $n$-th PU is less than the thresholds ${I_{c,th}}$ and ${I_{p,th}}$, respectively. To guarantee the QoS of each CU, the constraint (\ref{P1C7}) needs to be satisfied.

It is obvious that $({{\rm{P}}{\rm{1}}})$ is a non-convex multi-objective optimization problem due to the coupling of optimization variables and the fractional functions, so in the next section, a joint optimization algorithm is proposed to solve this multi-objective optimization problem.
\section{Joint Precoding And Common Stream Rate Optimization}
\subsection{Problem Reformulation} 
To solve this multi-objective non-convex optimization problem, the $\varepsilon $- constraint is applied in this paper, and the objective function of spectral efficiency is transformed into a constraint, then the problem $({{\rm{P}}{\rm{1}}})$ is reformulated in the following form
\begin{subequations}
		\setlength{\abovedisplayskip}{4pt}
	\setlength{\belowdisplayskip}{4pt}
	\begin{align}
		&({{\rm{P}}2}):{\rm{~}}\mathop {\max }\limits_{{\bf{P}},{\bf{c}}} {\rm{~}}{\eta _{EE}},\\
		&~~{\rm{s}}{\rm{.t}}{\rm{.}}~~~(\ref{P1C1})-\rm{(\ref{P1C7})},\label{P2C1}\\
		&~~~~~~~~~{\rm{}}{\eta _{SE}} \ge {\eta _{\rm{0}}} .\label{eitase}
	\end{align}
\end{subequations}
\begin{remark}
\emph{The initial selection of ${\eta _{\rm{0}}}$ should be less than ${\eta _{SE,\max }}$, which is independently obtained by\textup{\cite{Zarandi2020}} based on constraints \textup{(\ref{P1C1})-(\ref{P1C7})}. So the constraint \textup{(\ref{eitase})} guarantees that $\eta_{SE}$ is not less than $\eta_0$ in the process of solving all feasible solutions, which satisfies the condition required by Pareto optimal.}
\end{remark}
\subsection{Precoding Matrix and Common Stream Rate Joint Design} 
Let ${{\bf{Q}}_l} = {{\bf{p}}_l}{\bf{p}}_l^H$, and ${\bf{Q}} = \left[ {{{\bf{Q}}_1},{{\bf{Q}}_2}, \cdots ,{{\bf{Q}}_K}} \right]$ denotes the collections of all matrices. The precoding matrix optimization problem can be rewritten as
\begin{subequations}
	\setlength{\abovedisplayskip}{4pt}
	\setlength{\belowdisplayskip}{4pt}
\begin{align}
	&({{\rm{P}}{{\rm{2}}{\rm{.1}}}}):{\rm{}}~\mathop {\max }\limits_{{{\bf{Q}}}} ~{\rm{}}\frac{{\sum\nolimits_{k=1}^K {\left( {{C_k} + {R_{p,k}}} \right)} }}{{\sum\nolimits_l {{\rm{Tr}}\left( {{{\bf{Q}}_l}} \right)}  + {P_{cir}}}},\\
	&~~{\rm{s}}{\rm{.t}}{\rm{.}}~~~~~\rm{\left(\ref{Ck1}\right),\left(\ref{P1C7}\right),\left(\ref{eitase}\right)},\\
	&~~~~~~~~~~\sum\nolimits_l {{\rm{Tr}}\left( {{{\bf{Q}}_l}} \right)}  + {P_{cir}} \le {P_{\max }}\label{Tr}\\ 
	&~~~~~~~~~~~{{\bf{Q}}_l}\succeq0,\forall l, \label{semi}\\ 
	&~~~~~~~~~~~\rm{rank}\left( {{{\bf{Q}}_l}} \right) = 1,\forall l,\label{rank}\\
	&~~~~~~~~~~~{\rm{Tr}}\left( {{{\bf{F}}_n}{{\bf{Q}}_c}} \right) \le {I_{c,th}},\forall n,\label{Ic}\\
	&~~~~~~~~~~\sum\nolimits_{k=1}^K {{\rm{Tr}}\left( {{{\bf{F}}_n}{{\bf{Q}}_k}} \right)}  \le {I_{p,th}},\forall n,\label{Ip}
\end{align}
\end{subequations}
where ${{\bf{F}}_n}={{\bf{f}}_n}{{\bf{f}}_n}^H \in {\mathbb{H}^M},~{{\bf{f}}_n} = {\rm{diag}}\left( {\bf{h}} \right){{\bf{g}}_n}.$ 

Since the objective function is a fractional function, the optimal value of the parameter $\lambda  \in \mathbb{R}$, which can be given by Dinkelbach's algorithm\cite{DB}, is introduced in this paper. Let ${\tilde P_{{\rm{tot}}}}={\sum\nolimits_l {{\rm{Tr}}\left( {{{\bf{Q}}_l}} \right)}  + {P_{cir}}}$. Then, the objective function can be transformed into the following tractable form
\begin{equation}
	\setlength{\abovedisplayskip}{3pt}
	\setlength{\belowdisplayskip}{3pt}
	\begin{array}{c}
	{\tilde \eta _{EE}} = \sum\nolimits_{k=1}^K {\left[ {{C_k} + W\left( {{f_k}\left( {\bf{Q}} \right) - {v_k}{{\left( {\bf{Q}} \right)}}} \right)} \right]}- \lambda{\tilde P_{{\rm{tot}}}} .\label{obj}
	\end{array}
\end{equation}
 Where ${f_k}\left( {\bf{Q}} \right)$ and ${v_k}\left( {\bf{Q}} \right)$ are shown below
\begin{equation}
	\setlength{\abovedisplayskip}{3pt}
	\setlength{\belowdisplayskip}{3pt}
	{f_k}\left( {\bf{Q}} \right) = \log_2 \left( {\sum\nolimits_{i = 1}^K {{\rm{Tr}}\left( {{{\bf{F}}_k}{{\bf{Q}}_i}} \right)}  + {\sigma ^2}} \right),\forall k,
\end{equation}
and
\begin{equation}
	\setlength{\abovedisplayskip}{3pt}
	\setlength{\belowdisplayskip}{3pt}
	{v_k}\left( {\bf{Q}} \right) = \log_2 \left( {\sum\nolimits_{i \ne k}^K {{\rm{Tr}}\left( {{{\bf{F}}_k}{{\bf{Q}}_i}} \right)}  + {\sigma ^2}} \right),\forall k,
	\end{equation}
\setlength{\abovedisplayskip}{1pt}
\setlength{\belowdisplayskip}{3pt}
Since ${R_{p,k}}={W\left( {{f_k}\left( {\bf{Q}} \right) - {v_k}{{\left( {\bf{Q}} \right)}}} \right)}$ is the difference of two concave functions, it is a standard DC function. Therefore, a first-order Taylor expansion of ${v_k}\left( {\bf{Q}} \right)$ using successive convex approximations (SCA) is applied to transform Eq. (\ref{obj}) into a concave function, and the upper bound of ${v_k}\left( {\bf{Q}} \right)$ is as follows
\begin{equation}
	\setlength{\abovedisplayskip}{3pt}
	\setlength{\belowdisplayskip}{3pt}
\begin{array}{l}
	{v_k}{\left( {\bf{Q}} \right)^{ub}} \buildrel \Delta \over = {v_k}\left( {{{\bf{Q}}^{\left( r \right)}}} \right) + {\rm{vec}}{\left( {\nabla {v_k}\left( {{{\bf{Q}}^{\left( r \right)}}} \right)} \right)^T}{\rm{vec}}\left( {{\bf{Q}} - {{\bf{Q}}^{\left( r \right)}}} \right),
\end{array}
\end{equation}
where $\nabla {v_k}\left( {\bf{Q}} \right) = \left[ {\frac{{\partial {v_k}\left( {\bf{Q}} \right)}}{{\partial {{\bf{Q}}_1}}}, \cdots ,\frac{{\partial {v_k}\left( {\bf{Q}} \right)}}{{\partial {{\bf{Q}}_K}}}} \right]$ denotes the gradient of ${v_k}\left( {\bf{Q}} \right)$, ${\rm{vec}}$ denotes the vectorization operation, and ${{{\bf{Q}}^{\left( r \right)}}}$ represents the value of ${\bf{Q}}$ for the $r$-th iteration. The partial derivative of ${v_k}\left( {\bf{Q}} \right)$ with respect to ${{\bf{Q}}_i}$ can be expressed as
\begin{equation}
	\setlength{\abovedisplayskip}{4pt}
	\setlength{\belowdisplayskip}{3pt}
\frac{{\partial {v_k}\left( {\bf{Q}} \right)}}{{\partial {{\bf{Q}}_i}}} = \left\{ 
\begin{aligned}
		{\frac{{{{\bf{F}}_k}}}{{\ln 2\left( {\sum\nolimits_{i \ne k}^K {{\rm{Tr}}\left( {{{\bf{F}}_k}{{\bf{Q}}_i}} \right)}  + {\sigma ^2}} \right)}},{\rm{  o}}{\rm{.w,}}}\\
		{{\rm{}}{\bf{0}}{\rm{,}}~~~~~~~~~~~~~~~i = k.}
\end{aligned}
 \right.
\end{equation}
Thus the approximation of $R_{p,k}$ can be expressed as
\begin{equation}
	\setlength{\abovedisplayskip}{3pt}
	\setlength{\belowdisplayskip}{3pt}
	{\widetilde R_{p,k}} = W\left({f_k}\left( {\bf{Q}} \right) - {v_k}{\left( {\bf{Q}} \right)^{ub}}\right),\forall k,
\end{equation}
Based on the above approximation, all equations containing ${R_{p,k}}$ are transformed into concave functions or convex sets. It can be seen that the constraint (\ref{Ck1}) in problem $({{\rm{P}}{{\rm{2}}{\rm{.1}}}})$ is still a non-convex constraint, which can be transformed into the following form
\begin{equation}
	\setlength{\abovedisplayskip}{4pt}
	\setlength{\belowdisplayskip}{4pt}
	{\rm{Tr}}\left( {{{\bf{F}}_k}{{\bf{Q}}_c}} \right) \ge {\gamma _{c0}}\left( {\sum\nolimits_{i \ne k}^K {{\rm{Tr}}\left( {{{\bf{F}}_k}{{\bf{Q}}_i}} \right)}  + {\sigma ^2}} \right),\forall k,\label{gama}
\end{equation}
where ${\gamma _{c0}} = {2^{{R_c}/W}} - 1$ denotes the signal-to-noise ratio corresponding to the common stream rate $\sum\nolimits_{k = 1}^K {{C_k}}$.

Using the semidefinite relaxation (SDR) technique, the relaxation constraint (\ref{rank}) is relaxed, and in order to emphasize the relaxation constraint ${\rm{rank}}\left( {{{\bf{Q}}_l}} \right) = 1,\forall l,$ and at the same time to obtain the solution $\bf{P}^{*}$ of the problem $({{\rm{P}}{{\rm{2}}{\rm{.1}}}})$, the sequential rank one constraint relaxation technique (SROCR) is applied in this paper. This algorithm relaxes the rank-one constraint to an inequality by introducing auxiliary variables $\omega$ in the equivalent expression as follow 
\begin{equation}
	{\bf{u}_{\max }}{\left( {{\bf{Q}}_l^{\left( r \right)}} \right)^H}{{\bf{Q}}_l}{\bf{u}_{\max }}\left( {{\bf{Q}}_l^{\left( r \right)}} \right) \ge {\omega ^{\left( r \right)}}{\rm{Tr}}\left( {{{\bf{Q}}_l}} \right),\forall l,\label{rankone}
	\end{equation}
where $\bf{u}_{\max }$ denotes the eigenvector corresponding to the largest eigenvalue of ${{\bf{Q}}_l^{\left( r \right)}}$, and $r$ denotes the iteration number. Accoring to \cite{2017A}, the rank-one constraint can be satisfied gradually by updating $\omega$, which makes it easier to find a feasible solution.

Finally, the original problem $({{\rm{P}}{{\rm{2}}{\rm{.1}}}})$ can be written in the following solvable form
\begin{subequations}
	\setlength{\abovedisplayskip}{4pt}
	\setlength{\belowdisplayskip}{4pt}
	\begin{align}
		&({{\rm{P}}{{\rm{2}}{\rm{.2}}}}):{\rm{~}}\mathop {\max }\limits_{{\bf{Q}},{\bf{c}}} {\rm{~}} \sum\nolimits_{k = 1}^K {\left( {{C_k} + {\widetilde R_{p,k}}} \right)} 
		{\rm{}} - \lambda {\tilde P_{{\rm{tot}}}}\\ 
		&~~{\rm{s}}{\rm{.t}}{\rm{.}}~~~~~\rm{(\ref{Ck2}),(\ref{Tr}),(\ref{semi}),(\ref{Ic}),(\ref{Ip}),(\ref{gama}),(\ref{rankone}),}\\
		&~~~~~~~~~~~{\rm{}}{C_k} + {\widetilde R_{p,k}} \ge {R_{th}},\forall k,\\
		&~~~~~~~~~~{\rm{}}\sum\nolimits_{k = 1}^K {\left( {{C_k} + {\widetilde R_{p,k}}} \right)}  \ge W{\eta _{\rm{0}}}.\label{rate_tilde}
	\end{align}
\end{subequations}
The problem $({{\rm{P}}{{\rm{2}}{\rm{.2}}}})$\footnote{We utilize SCA to deal with non-convexity and obtain a suboptimal high-precision solution to the problem ({{\rm{P}}{{\rm{2}}{\rm{.2}}}}) by iteration, with convergence guaranteed by the interior point method\cite{CVX}.} is a joint convex problem on the variables $\bf{Q}$ and $\bf{c}$, and is a semi-definite programming (SDP) problem, which can be solved by using the CVX toolbox.

\subsection{Complexity Analysis of the Proposed Algorithm}
The proposed algorithm can be summarized as {\bf{Algorithm 1}}, and the complexity of the algorithm is mainly determined by its step 4. The complexity of the step is ${\cal O}\left( {\log \left( {1/{\varepsilon _0}} \right){{\left( M \right)}^{3.5}}} \right)$, and ${\varepsilon _0}$ is the accuracy of stopping iteration, which is set to ${\varepsilon _0} = {10^{ - 3}}$ in this paper.
\begin{algorithm}[htbp]
	\renewcommand{\algorithmicrequire}{\textbf{Input:}}
	\renewcommand{\algorithmicensure}{\textbf{Output:}}
	\caption{The Joint Optimization Algorithm}
	\label{alg1}
	\begin{algorithmic}[1]
		\STATE Solve ${\eta _{SE}}$ optimization problem to obtain ${\eta _{SE,max}}$.
		\STATE {\bf{Initialization}}: ${\bf{P}}^{0}$, ${\bf{c}}^{0}$, ${\eta _{\rm{0}}}$, ${\varepsilon _0}$, $\lambda_0$, $r=0$.
		\REPEAT
		\STATE Solve Problem $({{\rm{P}}{{\rm{2}}{\rm{.2}}}})$ and obtain ${\bf{c}}^{(r)}$ and ${\bf{P}}^{(r)}$ based on its solution using the SROCR technique.
		\STATE Get $\lambda$ using Dinkelbach's algorithm.
		\STATE $r \leftarrow r + 1$
		\UNTIL The fractional decrease of the objective value is
 		below a threshold ${\varepsilon _0}$.
        \STATE {\bf{return}} Precoding matrix and common stream rate vector.
	\end{algorithmic}  
\end{algorithm}

\section{Numerical Results}
In this section, the performance of the proposed algorithm is investigated. PBS serves $N=5$ PUs with a single antenna and CBS serves $K=5$ CUs with a single antenna. These users are randomly distributed in a radius circle with CBS as the coordinate origin, PUs are distributed in a radius of 500 m and CUs are distributed in a radius of 350 m. Unless specified, other simulation parameters are set as in Table \uppercase\expandafter{\romannumeral1}.
\begin{table}[H]
	\caption{Simulation Parameters}
	\begin{center}
		\resizebox{0.45\textwidth}{!}{
\begin{tabular}{l l c}  
	\toprule[1pt]
	\bf{Parameters} &~~~& \bf{Value}\\ 
	\midrule[0.5pt]
	Path-loss exponent ($\alpha $) &~~~~~~~~~~~~& 2.5  \\
	Rician factor (${\kappa _v}$) &~~~~~~~~~~~~& 1  \\
	Carrier frequency ($f$) &~~~~~~~~~~~~& 3 GHz \\
	System bandwidth ($W$) &~~~~~~~~~~~~& 20 MHz \\
	QoS of CUs (${R_{th}}$) &~~~~~~~~~~~~& 1 Mbps \\
	PUs common stream interference (${I_{c,th}}$) &~~~~~~~~~~~~& -80 dBm \\
	PUs private stream interference (${I_{p,th}}$) &~~~~~~~~~~~~& -60 dBm \\
	Noise power (${\sigma ^2}$) &~~~~~~~~~~~~& -90 dBm \\
	\bottomrule[1pt]
\end{tabular}
}
	\end{center}
\end{table}
\begin{figure*} [htbp]
	\centering
	\subfloat[\label{fig:a}]{
		\includegraphics[scale=0.40]{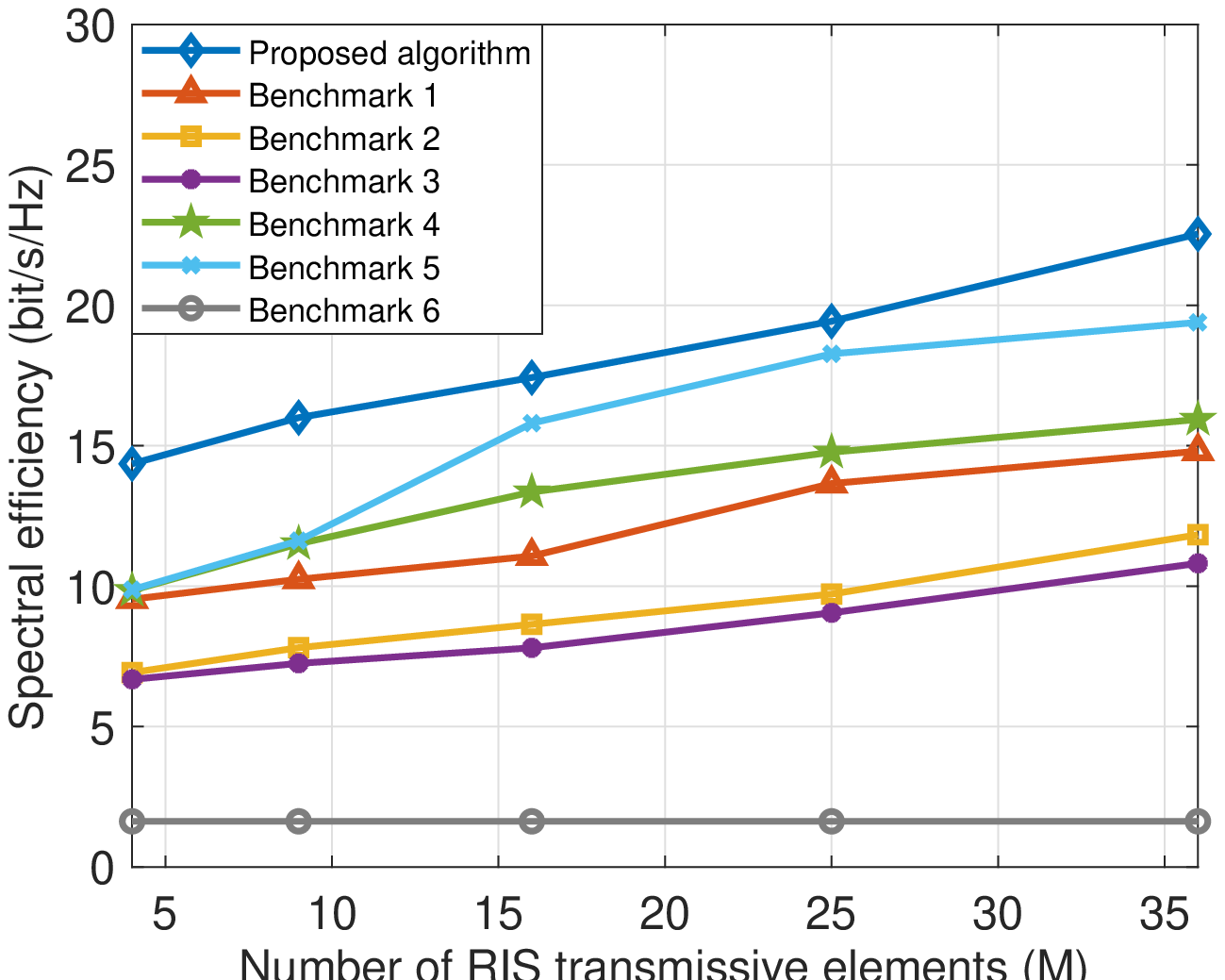}}
	\subfloat[\label{fig:b}]{
		\includegraphics[scale=0.40]{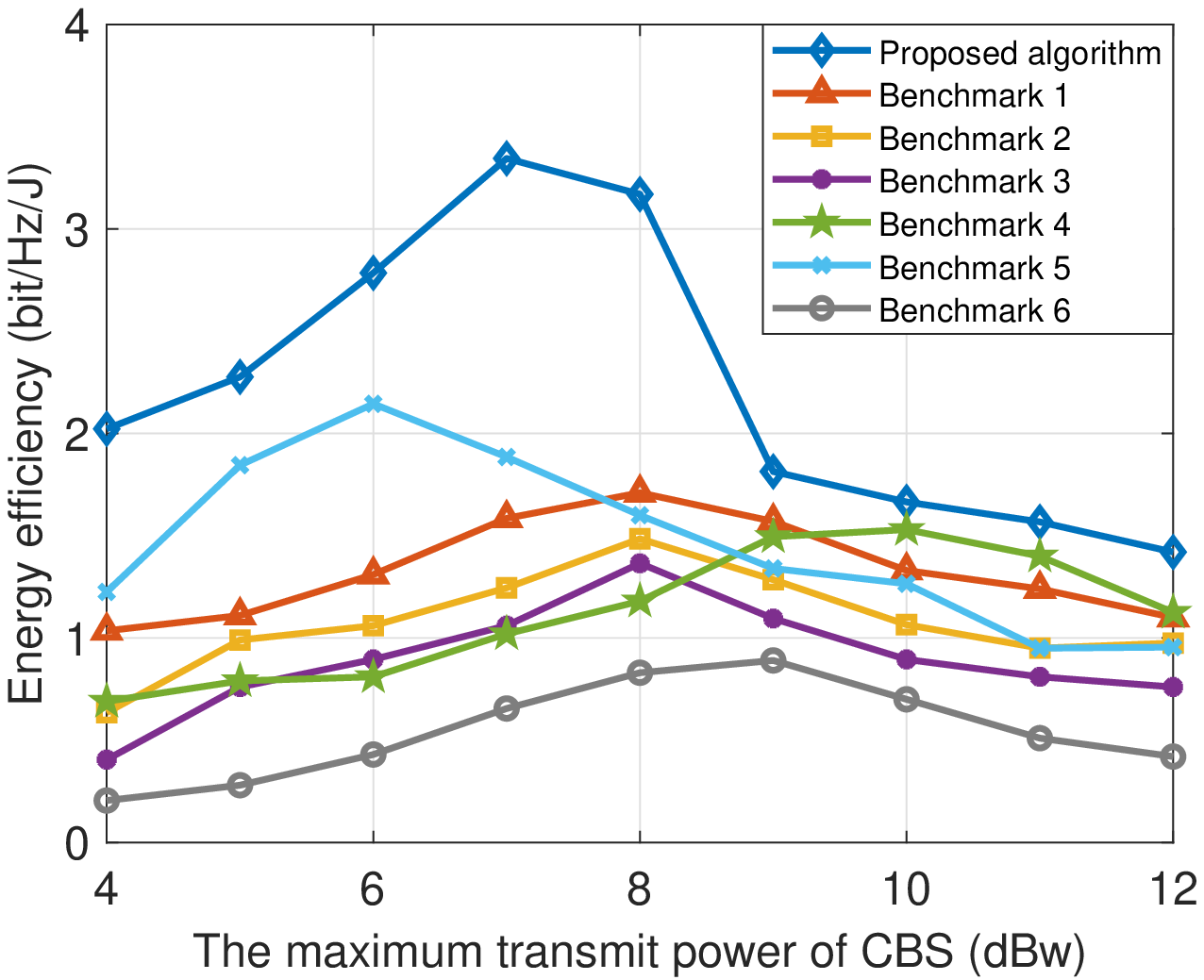}}
	\subfloat[\label{fig:c}]{
		\includegraphics[scale=0.40]{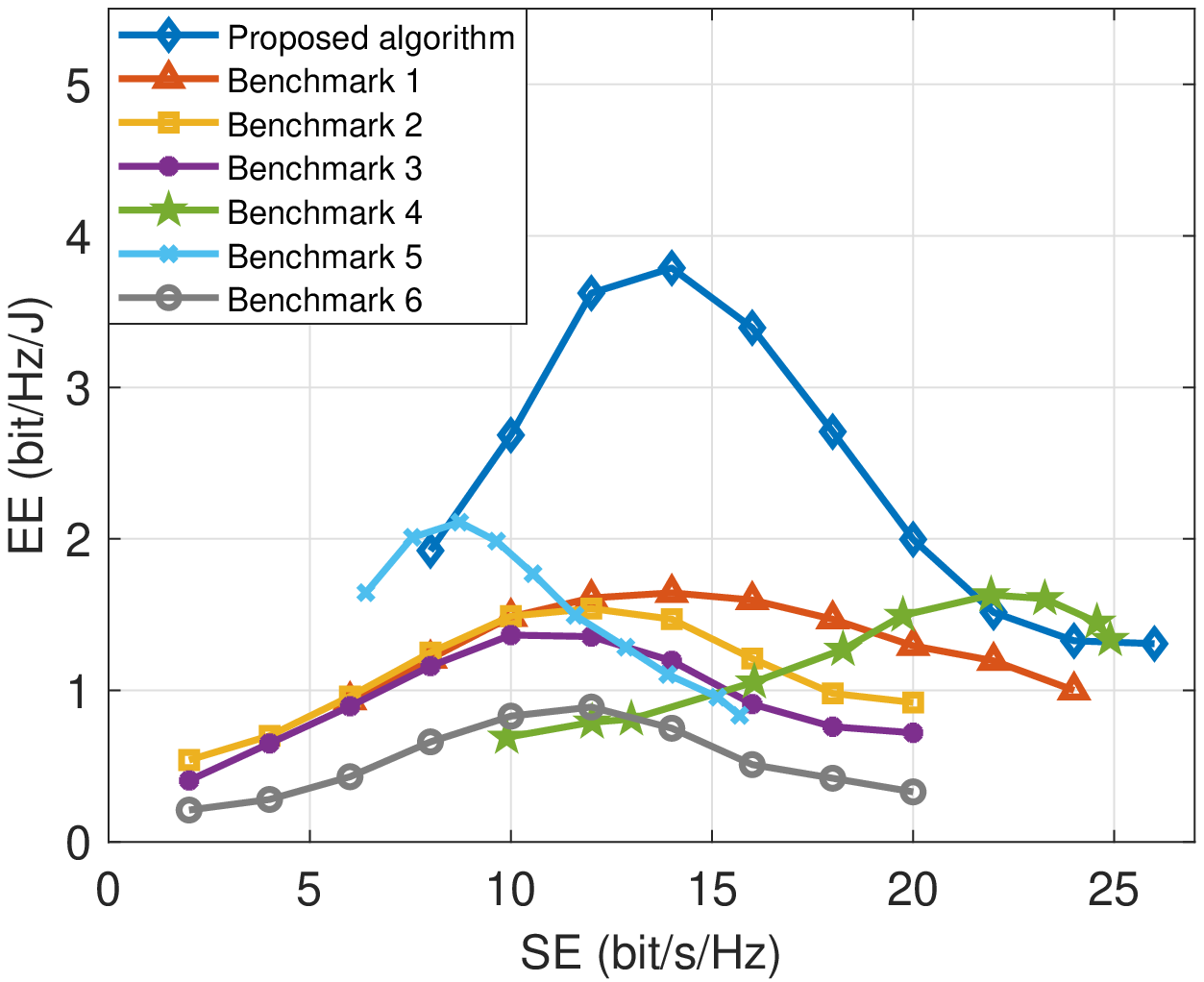}}
	\caption{Simulation Results: (a) SE versus the number of RIS transmissive elements (${P_{\max }}=10~{\rm{dBw}}, {\eta _{\rm{0}}}=0.9{\eta _{\rm{max}}}$). (b) EE versus the maximum power constraint of CBS ($M=9, {\eta _{\rm{0}}}=0.5{\eta _{\rm{max}}}$). (c) EE and SE trade-off under different CBS maximum transmit power ($M=9, {\eta _{\rm{0}}}=0.4{\eta _{\rm{max}}}$).}
	\label{fig2}
\end{figure*}

In this paper, we compare the performance of the proposed algorithm and other benchmarks as follows: (1) {\bf{benchmark 1}} (i.e., EE maximization): this scheme aims to
	maximize EE. (2) {\bf{benchmark 2}} (i.e., random precoding): this
	scheme does not optimize the precoding matrix and employs
	a randomly generated scheme. (3) {\bf{benchmark 3}} (i.e., fixed
	precoding): this scheme uses a fixed precoding matrix. (4) {\bf{benchmark 4}} (i.e., SDMA): users access by SDMA. (5) {\bf{benchmark 5}} (i.e., NOMA): users access by NOMA. (6) {\bf{benchmark 6}} (i.e., No RIS): this scheme does not deploy
	a TRIS.

We first provide insight into the relationship between the SE of the system and the number of TRIS elements. As shown in Fig. \ref{fig2}\subref{fig:a}, it is obvious that the SE of all schemes improves with the increase in the number of TRIS elements, except for benchmark 6 where RIS is not deployed, and despite taking random and fixed schemes, benchmark 2 and 3 still have higher SE than benchmark 6, proving that TRIS has a significant effect on enhancing the SE of the system. Meanwhile, the SE in the proposed architecture is higher than SDMA, NOMA, and No RIS, confirming that the combination of RSMA and TRIS can bring higher gains.

Then, we shed light on the variation of the EE with the maximum power constraint of CBS. As shown in Fig. \ref{fig2}\subref{fig:b}, due to the change in the growth rate of the $R_{tot}$ and the $P_{tot}$, there are intersections of the two functions, resulting in a rising and then falling trend of EE with the variation of $P_{tot}$. It is noteworthy that the architecture in this paper can achieve higher EE with lower $P_{tot}$ compared to other schemes, which confirmes that the proposed architecture can achieve high-rate multi-stream communication with low power consumption.

Finally we investigate the tradeoff between EE and SE as the maximum power constraint of CBS is increased from 4 dBw to 13 dBw. Fig. \ref{fig2}\subref{fig:c} shows that EE increases and then decreases as SE increases, which indicates that higher SE requires more energy consumption, and the rate of increase of SE is less than the rate of increase of energy consumption, leading to a decrease in EE. As the total power goes from 11dBw to 13dBw, the common stream power of RSMA is close to the private stream power, resulting in the degradation of RSMA to SDMA, so their EEs are neck and neck. Meanwhile, the proposed algorithm achieves higher SE and EE and has a better SE and EE balance than other schemes, confirming the necessity and effectiveness of the proposed algorithm to jointly optimize SE and EE.

\section{Conclusions}
In this paper, we proposed TRIS empowered cognitive RSMA networks to implement low-power, high-rate multi-stream communication. Based on this networks, the joint precoding matrix and common stream rate design optimization problem has been formulated. To solve the formulated non-convex problem, we utilized a joint optimization algorithm framework based on DC programming and SCA to obtain the solution. Based on the experimental results, we provided design guidelines to improve the SE of the system by increasing the number of low-cost TRIS elements, and the transmitter transmit power is maintained at an appropriate level to ensure maximum EE.
\bibliographystyle{IEEEtran}
\bibliography{IEEEabrv,reference}

%
%
%
%
%
%
%
%

\end{document}